\documentclass[
aps,
prl,
preprint,
floatfix,
superscriptaddress,
showkeys,
amsmath,
amssymb,
longbibliography,
floatfix]{revtex4-2}
\usepackage[english]{babel}
\usepackage{amsmath,amssymb,bbm,mathrsfs,bm,braket,color,graphicx,comment,amsfonts,dsfont}
\usepackage[colorlinks,linkcolor=blue,citecolor=blue,urlcolor=blue]{hyperref}
\usepackage[mathscr]{euscript}
\usepackage{physics}
\usepackage{xcolor}
\usepackage{ulem}
\usepackage{bm}
\usepackage{orcidlink}
\usepackage{multirow}

\def\beq{\begin{equation}}
\def\eeq{\end{equation}}

\begin{document}

\title{Probing the shape of the Weyl Fermi surface of NbP using transverse electron focusing} 
\author{F. Balduini} \email{ico@zurich.ibm.com} \affiliation{IBM Research - Zurich, 8803 Ruschlikon, Switzerland}
\author{L. Rocchino}\affiliation{IBM Research - Zurich, 8803 Ruschlikon, Switzerland}
\author{A. Molinari}\affiliation{IBM Research - Zurich, 8803 Ruschlikon, Switzerland}
\author{T. Paul} \affiliation{IBM Research - Zurich, 8803 Ruschlikon, Switzerland}
\author{G. Mariani}\affiliation{IBM Research - Zurich, 8803 Ruschlikon, Switzerland}
\author{V. Hasse}\affiliation{ Max Planck Institute for Chemical Physics of Solids, 01187 Dresden, Germany}
\author{C. Felser }\affiliation{ Max Planck Institute for Chemical Physics of Solids, 01187 Dresden, Germany}
\author{C. Zota}\affiliation{IBM Research - Zurich, 8803 Ruschlikon, Switzerland}
\author{H. Schmid}\affiliation{IBM Research - Zurich, 8803 Ruschlikon, Switzerland}
\author{B. Gotsmann} \email{bgo@zurich.ibm.com} \affiliation{IBM Research - Zurich, 8803 Ruschlikon, Switzerland}

\date{\today}
             
\maketitle

\textbf{
The topology of the Fermi surface significantly influences the transport properties of a material. Firstly measured through quantum oscillation experiments, the Fermi surfaces of crystals are now commonly characterized using angle-resolved photoemission spectroscopy (ARPES), given the larger information volume it provides.
In the case of Weyl semimetals, ARPES has proven as a remarkably successful method in verifying the existence of the Weyl points and the Fermi arcs, which define a Weyl Fermi surface.
However, ARPES is limited in resolution, leading to significant uncertainty when measuring relevant features such as the distance between the Weyl points. While quantum oscillation measurements offer higher resolution, they do not reveal insights into the cross-sectional shape of a Fermi surface.
Moreover, both techniques lack in providing critical information about transport, like the carriers mean free path. 
Here, we report measurements unveiling the distinctive peanut-shaped cross-section of the Fermi surface of Weyl fermions and accurately determine the separation between Weyl points in the Weyl semimetal NbP. 
To surpass the resolution of ARPES, we combine quantum oscillation measurements with transverse electron focusing (TEF) experiments, conducted on microstructured single-crystals.  
The TEF spectrum relates to the Fermi surface shape, while the frequency of the quantum oscillations to its area. Together, these techniques offer complementary information, enabling the reconstruction of the distinctive Fermi surface geometry of NbP, which originates from the combination of two circular Fermi surfaces filled by carriers of opposite chirality. 
Concurrently, we extract the electrical transport properties of the bulk Weyl fermions.
Our work showcases the integration of quantum oscillations and transverse electron focusing in a singular experiment, allowing for the measurements of complex Fermi surface geometries in high-mobility quantum materials.
}

Weyl semimetals (WSMs) are topological semimetals where linearly dispersive conduction and valence bands cross at discrete points in the energy-momentum space, called Weyl points. The Weyl points always come in pairs of opposite chirality, as they originate from the splitting of a degenerate Dirac point when either time-reversal or space-inversion symmetry is broken. Topological states called Fermi arcs connect the Weyl points at the surface \cite{armitage_weyl_2018}. 
WSMs have been discovered using angle-resolved photoemission spectroscopy (ARPES) \cite{xu_discovery_2015}, which allows to measure the distinctive features of the Weyl Fermi surface, namely the Weyl cones, Weyl points, and Fermi arcs. However, due to the inherently limited resolution of ARPES, particularly in the context of the typically small Fermi surfaces of semimetals, it fails to accurately determine the shape of the Weyl Fermi surface and quantify its dimension \cite{armitage_weyl_2018}.
In this context, quantum oscillations have emerged as a powerful tool for investigating topological semimetals in particular, providing a means to measure the Fermi surface dimension with significantly enhanced resolution compared to ARPES, and providing information about its topological character \cite{alexandradinata_fermiology_2023, guo_temperature_2021}.
Still, quantum oscillations don't carry information about the shape of the cross-section of the Fermi surface. This means that no technique precisely measures the Fermi surface topology of materials, which is critical in determining properties such as the magnetic response \cite{zhang_magnetoresistance_2019}, the transport directionality \cite{bachmann_directional_2022}, or the emergence of a charge density wave \cite{rice_new_1975}. Specifically, in the case of WSMs, a precise measurement of the distance between a pair of Weyl nodes is missing. This distance, determined by spin-orbit coupling, defines the length of the Fermi arcs \cite{liu_evolution_2016}.

To address this constraint, we use a synergistic approach that combines quantum oscillations with transverse electron focusing (TEF) to infer the shape of the Fermi surface of the Weyl electrons in NbP, and precisely measure its dimensions. 
In a TEF experiment, a pair of micro-nozzles injects and probes charged carriers undergoing cyclotron motion. Given that the cyclotron orbits in a solid mimic the contour of the Fermi surface, the spectrum obtained from a TEF experiment depends on the Fermi surface topology, and the position of the peaks on the main lengths of the Fermi surface \cite{ bachmann_super-geometric_2019, tsoi_studying_1999, heil_electron_2000}. Furthermore, information regarding the scattering mechanism and mean free path can be extracted \cite{gupta_precision_2021, tsoi_studying_1999}. 
On the other hand, the frequency of the quantum oscillations is proportional to the cross-sectional area of the Fermi surface, offering additional insights alongside those given by a TEF experiment, thereby facilitating the construction of complex Fermi surface shapes. By varying the angle between the magnetic field and crystalline direction in a quantum oscillation experiment, one can determine the Fermi surface volume and thus the carrier density associated with a specific pocket of carriers \cite{shoenberg_magnetic_1984}. This complements the measurement of the mean free path provided by a TEF experiment to infer the electrical transport properties of a specific Fermi surface pocket.

Our data indicate that in NbP the Fermi level lies above the Weyl cones intersection line, which results in an achiral peanut-shaped Fermi surface originating from the merging of the two Weyl cones, as described by the Weyl Hamiltonian. 
Through quantification, we extract the distance between Weyl nodes in NbP, surpassing the precision achievable through ARPES measurements.
Moreover, the estimated transport properties suggest that the high mobility observed in NbP \cite{ shekhar_extremely_2015, balduini_origin_2024} may be attributed to the Weyl electrons, hosted in the achiral Fermi surface. In this regard, the TEF measurement surpasses the limitation of Hall and resistivity measurements which cannot discriminate the contribution of different types of carriers, such as trivial and Weyl carriers, in a multi-band material like NbP.\\

For our experiments, we start with single crystals of NbP grown using the physical vapor transport (see Methods) to have a low defect density. 
A well-defined contact geometry and arrangement is beneficial for high-quality measurement data. To this end, we prepared an NbP plate with seven microcontacts using focused ion beam (FIB) microstructuring \cite{moll_focused_2018}, as illustrated in Figure 1 b,c (the c-axes is out-of-plane). 
The schematic in Figure 1a shows the conventional configuration for TEF experiments: charge carriers are injected from one microcontact, the emitter E, and an external magnetic field $B$ is tuned to focus the carriers in the proximity of another microcontact, the collector C, where the voltage is measured. 
A peak in the collector's voltage as a function of field appears when the focusing condition is met:
\begin{equation} \label{focusing}
    B_{n,j} = \frac{2n\hbar k_{F_j}}{eL}
\end{equation}
where $n$ is an integer due to possible multiple specular reflections at the surface, $\hbar k_{F_j}$ is the j-th Fermi momentum of the electrons ejected parallel to the emitter, $e$ is the electron charge, and $L$ is the E-C distance.

As shown in Figure 1d, for temperatures lower than 100\,K and magnetic field lower than 1\,T, two prominent peaks appear at the collector's voltage, ascribable to the focusing of negatively charged ballistic carriers. 
Commonly, in  TEF experiments, signals periodic in field are present due to specular reflection of the carriers at the boundaries \cite{tsoi_studying_1999}. Here we do not observe any periodic signal, but this is not surprising considering the surface damage induced by FIB microstructuring, which causes the formation of a tens of nanometers thick amorphous layer \cite{kato_side-wall_1999, bachmann_inducing_2017}, likely favoring purely diffusive scattering at the boundaries. 
As the temperature is lowered below 25\,K and the field is increased above 1\,T, a transition occurs from real space effects to reciprocal space quantization, leading to the predominance of quantum oscillations in the collector signal.

The position of the peaks in Figure 1d carries valuable information about the Fermi surface shape and dimension. 
In Figure 2a, the position of the TEF peaks, which varies with the distance between the emitter and collector, is used to find the momenta of the electrons through a fit of Equation 1. The fit confirms the proportionality $L\sim B_j^{-1}$, and increases the accuracy of the extraction of the momenta: $k_{F1}$ = 0.022 $\mathrm{\AA^{-1}}$ and $k_{F2}$ = 0.035 $\mathrm{\AA^{-1}}$. 
In Figure 2c, despite the presence of two peaks in the TEF experiment associated with two distinct Fermi momenta, the quantum oscillations are dominated by a single frequency  $f_{\alpha}$ = 31 T (other frequencies exist, but they have much less amplitude). This frequency corresponds to a singular Fermi surface area, as per the Onsager relation $A_{\rm FS} = ( 2\pi^2 / \Phi_0) f_{\alpha} = 2.9\cdot10^{-3}\mathrm{\AA^{-2}} $, where $ \Phi_0 = 2.07 \cdot 10^{-15}$ $\mathrm{T m^2}$ is the flux quantum. 

Concurrently, the TEF and quantum oscillation amplitudes can be used to extract valuable information about the transport properties.
The mean free path $l_{mfp}$ associated with each TEF peak is estimated by fitting their amplitude as a function of the traveled distance $s(L)$, according to $ A_{TEF} = A_0 g(b,L) e^{-s(L)/l_{mfp}}$, where $b$ is the width of the contacts and $g(b,L)$ is a function that depends on the Fermi surface geometry \cite{bachmann_super-geometric_2019, tsoi_studying_1999} (Figure 2b).
In Figure 2d, by fitting the temperature dependence of the amplitude $A_{SdH}$ of the quantum oscillations using the Lifshitz-Kosevich formula $A_{SdH}=\frac{\lambda}{\sinh(\lambda)}$, where $\lambda = 14.7 \frac{T}{B m^*}$ and $T$ is the absolute temperature, we find the effective mass of the corresponding carriers $m^*$ = 0.13 $\mathrm{m_e}$, where $\mathrm{m_e}$ is the electron mass. 

The Fermi surface area and mass extracted from the quantum oscillations are consistent with the ones calculated using density functional theory (DFT) for the pocket hosting Weyl electrons in NbP ($(A_{\rm FS})_{th} = 2.8\cdot10^{-3}$ $\mathrm{\AA^{-2}}$, $(m^*)_{th}$ = 0.10 $\mathrm{m_e}$ \cite{lee_fermi_2015, klotz_quantum_2016}).
However, the radius of such Fermi surface area, if assumed to be circular, does not match any of the two momenta found from the TEF experiment, albeit closely aligning with their average. To interpret the TEF data accounting for this inconsistency, we will use a more realistic model for the cross-sectional shape of the Weyl Fermi surface, beyond the circular approximation.

To obtain the cross-section of a Weyl Fermi surface, we start with the low-energy Hamiltonian:
\begin{equation}
    H = v\tau (\bf{\sigma} \cdot \bf{k}) + m\tau_z + b\sigma_z
\end{equation}
which represent Weyl semimetals for $|b|>|m|$, where $v$, $m$, and $b$, are velocity, mass, and Zeeman-like parameters respectively \cite{armitage_weyl_2018}. The isoenergy cuts in the $k_z-k_y$ plane define the Fermi surface shape relevant to this experiment (see Figure 3a). When the Fermi level is above the Weyl cone's intersection, the Fermi surface manifests as a degenerate peanut-shaped structure.
In a TEF experiment, the real space cyclotron orbits traveled by the charge carriers exiting the emitter resemble the Fermi surface shape, rotated by 90 degrees and scaled by $\hbar/eB$.
Therefore, having the Fermi surface contour, we can calculate the probability of an electron reaching the collector to simulate the TEF spectrum, as shown in Figure 3b (more details can be found in the Supplementary information). 
Interestingly, by lowering the Fermi level below the Weyl cones' intersection, the TEF experiment would result in a qualitatively different spectrum, as demonstrated by the comparison of Figures 3b and 3c. This is because the Weyl Fermi surface would evolve from a singular peanut-like shape to two separated rounded Fermi surfaces, undergoing a Lifshitz transition.
In particular, the spectrum generated by a peanut-shaped Fermi surface in Figure 3b, characterized by two prominent peaks and a smaller one at lower fields, closely resembles the experimental data shown in Figure 3d, supporting that Weyl electrons in NbP occupy a Fermi surface with net zero chirality and Chern number.

The essential features of such Fermi surface are simplified to two fittable parameters: two rounded heads of radius $k_{F1}$ that form a body of length $2\times k_{F2}$, which correspond to the momenta $k_{F1}$ and $k_{F2}$ extracted in Figure 2a.
This allows us to compute the area of the Fermi surface of Weyl electrons in NbP from the TEF data, which we find to be $A_{\mathrm{FS}}$ =  $\mathrm{2.8\cdot10^{-3}}$ $\mathrm{\AA^{-2}}$. This value aligns well with the $2.9 \cdot 10^{-3}$ $\mathrm{\AA^{-2}}$ determined through the quantum oscillation experiment.

To further test the hypothesis that the two prominent peaks in the TEF signal originate from the Weyl Fermi surface, we used the FIB to etch out a cavity in the NbP sample, at a distance that would block a portion of the carriers that generate the peak at $B = B_1$, while letting through all the carriers belonging to $B = B_2$ and $B = B_0$ \cite{bachmann_manipulating_2020, aidala_imaging_2007} (Figure 3e). This manipulation resulted in a reduction of the TEF peak at B$_1$, while the peaks at B$_2$ and B$_0$ remained unchanged in both shape and magnitude, as expected according to the predicted peanut-shaped orbits. If the three peaks stem from three separated circular Fermi surfaces, they would have been affected in a similar manner.

Notably, NbP is the material with the smallest spin-orbit coupling within the archetypal Weyl semimetal family (including TaAs, TaP, NbAs, and NbP) \cite{liu_evolution_2016}. Consequently, it possesses the shortest distance between Weyl nodes and Fermi arc length. Traditional ARPES methods are often inadequate for resolving the node separation in NbP \cite{xu_observation_2015, souma_direct_2016, belopolski_observation_2015}. Even with high-resolution ARPES (HR-ARPES), measurements yield results with substantial uncertainty \cite{liu_evolution_2016}. By employing TEF, we achieve precise measurements of the separation between the Weyl nodes in NbP, $\Delta k_W = 2(k_{F2} - k_{F1}) = (2.53 \pm 0.07)\cdot10^{-2}$ $\mathrm{\AA^{-1}}$, significantly enhancing the current resolution of this quantity. A comparison of the resolution achievable with HR-ARPES and TEF methods to determine the Weyl nodes separation in k-space is shown in Figure 3g, along with the results of DFT calculation using the software packages Open-MX \cite{lee_fermi_2015} and VASP \cite{sun_topological_2015}. \\

Having identified the signals in quantum oscillations and TEF experiments linked to the Weyl electrons in NbP, we can study their transport properties.
Angle-dependent Shubniko-de Haas (SdH) oscillations, measured on a micro Hall-bar extracted from the same NbP single-crystal (Supplementary Information), are used to calculate the Fermi surface volume and associated carrier density, while the temperature-dependent amplitude to calculate the effective mass. 
The focusing fields of the TEF experiment are used to extract the Fermi momenta $k_F$, and their amplitudes decay, as the distance traveled by the electrons increases, are fitted to determine the mean free path $l_{mfp}$ (Figure 2d, fit of  $ A_{TEF} = A_0 g(b,L) e^{-s(L)/l_{mfp}}$ for $g(b,L)=\sqrt{b/L}$ and $s(L)$ calculated according to the Fermi surface geometry). The information extracted from the TEF data allows calculating the mobilities of the bulk Weyl electrons: $\mu = \frac{el_{mfp}}{\hbar k_F}$. We find $\mu_{We1} = 2.2 \cdot 10^5$ $\mathrm{cm^2/Vs}$ and $\mu_{We2} = 2.9 \cdot 10^5$ $\mathrm{cm^2/Vs}$. 
Finally, by combining the mobility extracted from TEF with the carrier density and effective mass from SdH, we can calculate the scattering time and conductivity of Weyl electrons,  $\tau = \frac{\mu m^*}{e}=$18 ps and $\sigma = ne\mu=$2.4$\cdot$10$^7$ S/m respectively. The calculated conductivity of Weyl electrons aligns with the low-temperature conductivity of the NbP Hall bar (Supplementary Information). This result indicates that the Weyl electrons are the primary contributors to the electrical conductivity of NbP at low temperatures, owing to their remarkably high mobility. We highlight that these are bulk electrons that occupy an achiral Fermi surface.\\

In summary, our study proposes the combination of transverse electron focusing and quantum oscillation experiments to probe the Fermi surface of high-mobility quantum materials with complex Fermi surface shapes. 
To demonstrate, we measured the Weyl Fermi surface of the Weyl semimetal NbP, unveiling its distinct peanut-shaped geometry resulting from the merging of two Weyl cones. This approach, allowed us to measure the separation between Weyl nodes with higher precision than has previously been achieved with ARPES.
In addition, from the same measurements, we also extracted the transport properties of the Weyl electrons contained in the bulk and achiral Weyl Fermi surface. Notably, we observed that these carriers are primarily responsible for NbP's remarkably high mobility, overcoming the limitations of conventional techniques such as Hall effect and resistivity measurements, which can't discriminate between carriers belonging to distinct Fermi surfaces. 
Our study on NbP demonstrates a way to characterize high mobility quantum materials with complex Fermi surfaces, providing crucial insights into their Fermi surface shapes, symmetries, and carriers type and transport properties.

\section{Methods}
\textbf{Crystal growth} \\
High-quality single bulk crystals of NbP were grown via a chemical vapor transport reaction using an iodine transport agent. A polycrystalline powder of NbP was synthesized by direct reaction of niobium (Chempur 99.9$\%$) and red phosphorus (Heraeus 9.999$\%$) within an evacuated fused silica tube for 48 h at 800 °C. The growth of bulk single crystals of NbP was then initialized from this powder by chemical vapor transport in a temperature gradient, starting from 850 °C (source) to 950 °C (sink), and a transport agent with a concentration of 13.5 mg cm$^{-3}$ iodine (Alfa Aesar 99.998$\%$). \\

\textbf{Sample preparation}\\
Microscopic bars were extracted from a NbP single crystal by means of focused ion beam (FIB) microstructuring \cite{moll_focused_2018}, which allows for high aspect-ratio samples with good control of geometry and crystalline direction, and homogeneous magnetic field distribution along the sample. As a drawback, the properties of a thin superficial layer are altered \cite{bachmann_inducing_2017}, nevertheless, bulk properties are unchanged as demonstrated by the good match between quantum oscillations in microstructured and bulk samples. 
After the milling procedure, the sample was placed on a silicon chip with a silicon oxide spacer and pattered gold lines, and contacted using ion-assisted chemical vapor deposition of Platinum (contacts resistance around 15 $\mathrm{\Omega}$). 
The sample, shown in Fig. 1b and 1c, consists of a  25 $\mathrm{\mu m}$ by  25 $\mathrm{\mu m}$ plate, 1 $\mathrm{\mu m}$ thick with a total of seven 250 nm-wide contacts not-equally spaced. This allows for 21 values of $L$, from 1.23 $\mathrm{ \mu m}$ to 11.18 $\mathrm{ \mu m}$, depending on the chosen E-C couple. However, signals for L $>$ 8 $\mathrm{\mu m}$ (electron trajectory s $>$ 25 $\mathrm{\mu m}$) were not clearly discernible from the noise, and were excluded from the analysis. \\

\textbf{Measurements}\\
Electrical measurements were performed in a cryostat (Dynacool from Quantum Design) using external lock-in amplifiers (MFLI from Zurich Instruments). An AC current of constant amplitude of around 50 $\mathrm{\mu A}$ at 113 Hz is used for both the SdH and TEF experiments.\\

\begin{acknowledgments}
We wish to acknowledge the support of the Cleanroom Operations Team of the Binnng and Rohrer Nanotechnology Center (BRNC). F.B., B.G. and T.P. acknowledge the SNSF open project HYDRONICS. under the Sinergia gran (no. 189924). L.R. and C.Z. acknowledge the SNSF Ambizione programme. A.M. acknowledges funding support from the European Union's Horizon 2020 research and innovation programme under the Marie Sklodowska-Curie grant agreement no. 898113 (InNaTo). C.Z., H.S., A.M., V.S, C.F., and B.G. acknowledge the FET open project no. 829044 (SCHINES). We thank Fabrizio Nichele, Thomas Ihn, and Philip Moll for the fruitful discussions.
\end{acknowledgments}


%

\newpage

\begin{figure*}
\includegraphics[width=\linewidth]{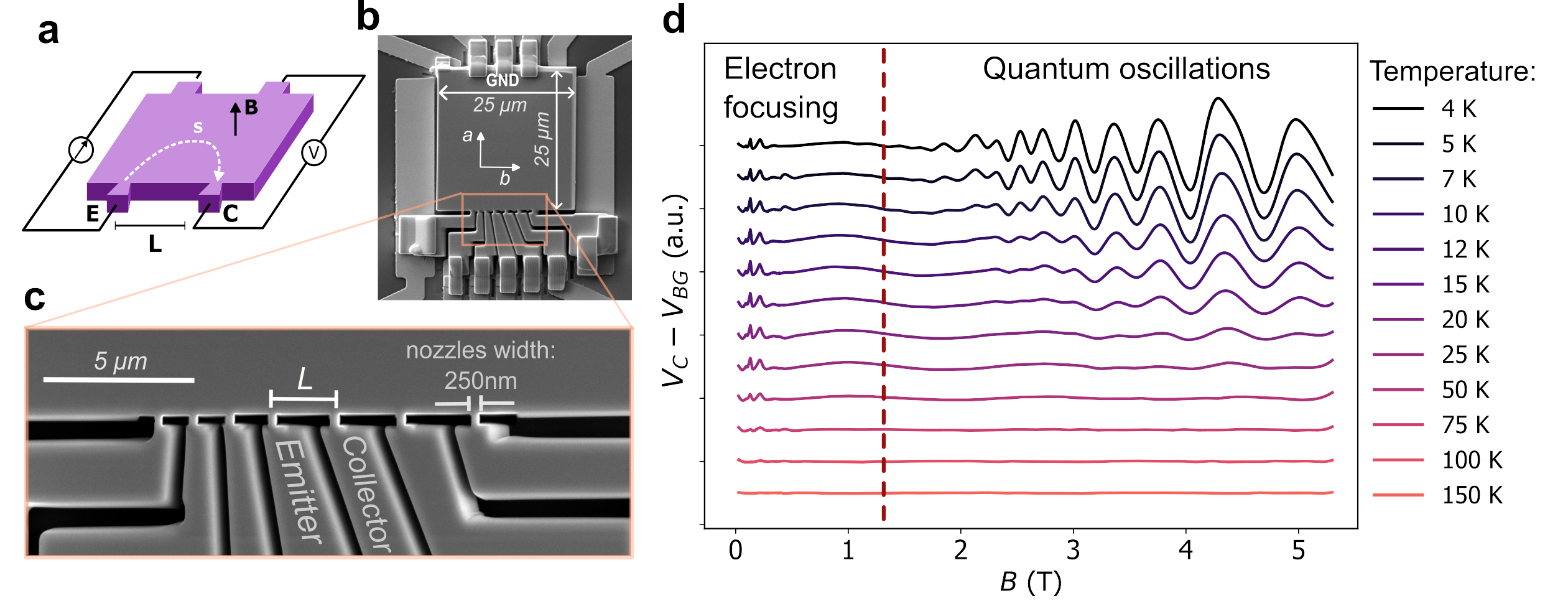}%
\caption{\textbf{Transverse electron focusing in NbP.} a) Schematic of a TEF experiment: charge carriers are ejected from an emitter E and are focused to a collector C, whose voltage is measured while sweeping the magnetic field B. The trajectory $s$ depends on the Fermi surface topology. b) Scanning electron micrograph of the NbP sample, oriented to have the charge carriers moving in the a-b plane. c) Close view of the contacts, here labeled for the emitter and collector electrodes used for the experiment in d. d) TEF data for an E-C distance of 2.2\,$\mu$m. The plot shows the collector signal minus a polynomial background as a function of the magnetic field and for temperatures ranging from 150 K to 4 K. At low fields, the collector signal is dominated by the transverse focusing of charged carriers; at high fields, quantum oscillations emerge.}
\end{figure*}

\begin{figure*}
\includegraphics[width=\linewidth]{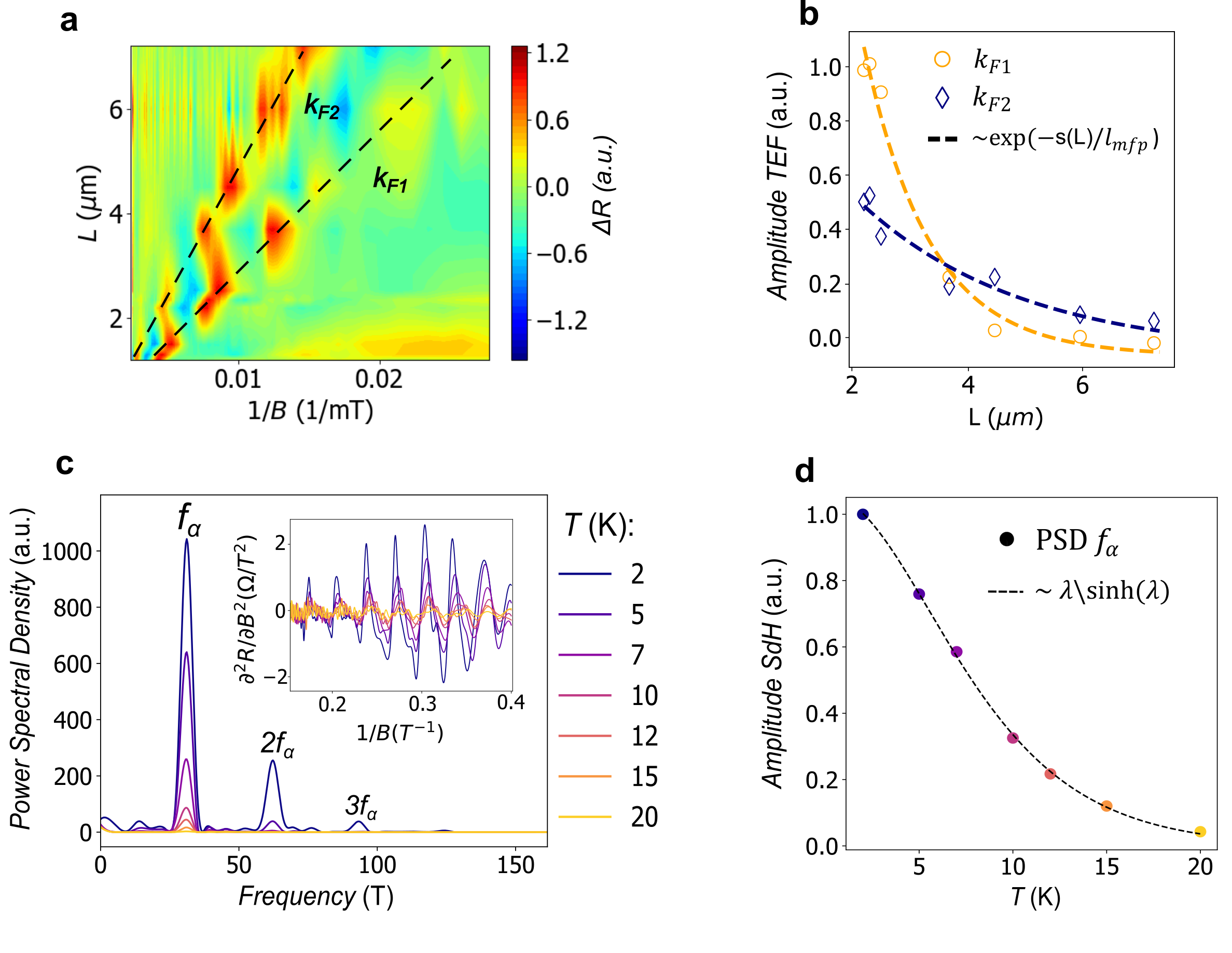}%
\caption{\textbf{Probing the Fermi surface using TEF and quantum oscillations.} a) 2D plot of the TEF signal for seven E-C distances $L$, after background removal and signal renormalization. A linear fit of the position of the peaks allows extracting the Fermi momenta of the electrons ejected parallel to the emitter. b) Amplitudes of the TEF peaks after background removal, and exponential decay fit to extract the mean free path $l_{mfp}$. c) Power spectral density (PSD) of the quantum oscillations showed in the inset, isolated performing a second derivative of the magnetoresistive data. d) The amplitude of the PSD of $f_\alpha$ is fitted as a function of the temperature to extract the effective mass of the carriers using the Livshits-Kosevitch equation.}
\end{figure*}

\begin{figure*}
\includegraphics[width=\linewidth]{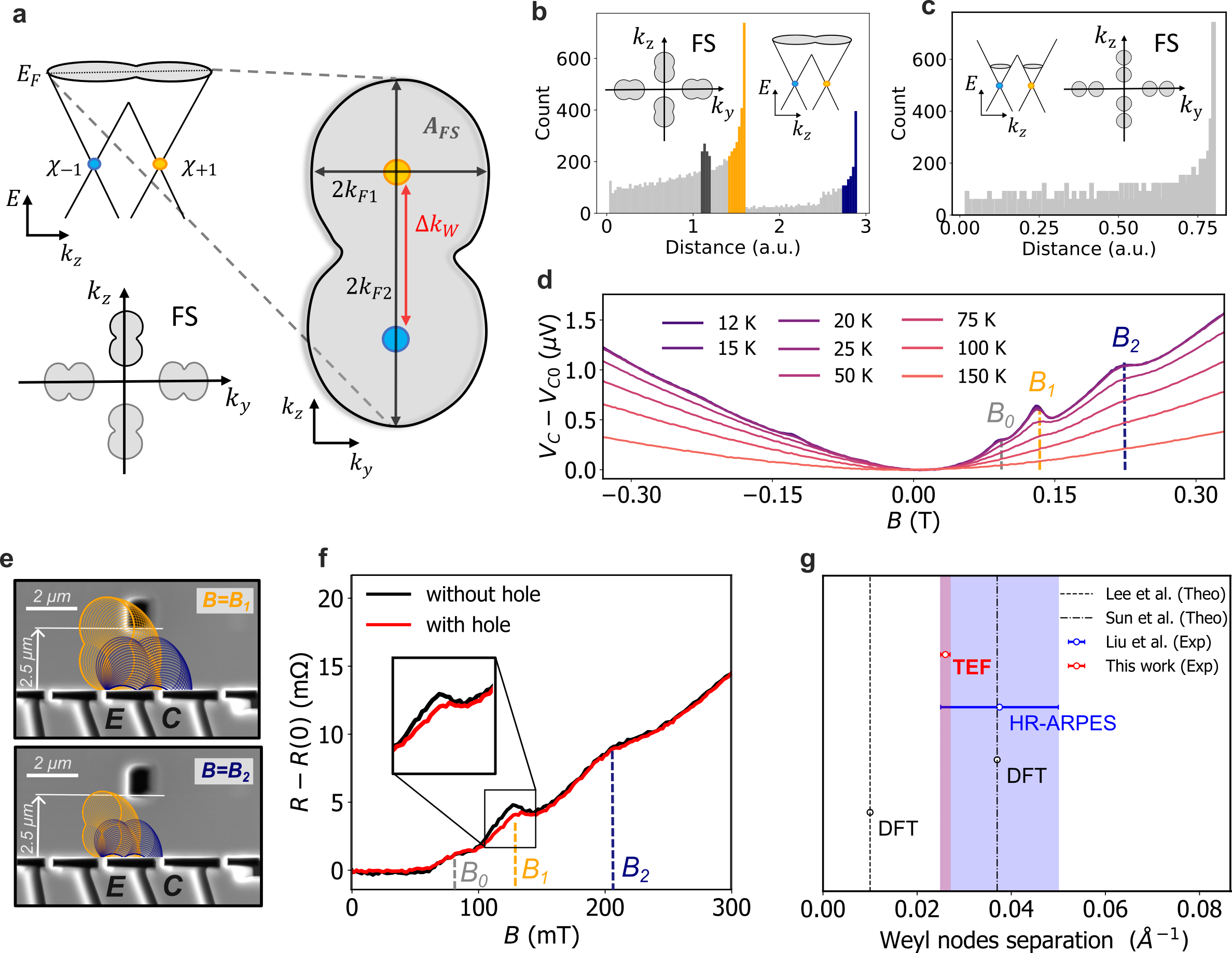}%
\caption{\textbf{Weyl Fermi surface reconstruction.} a) Schematic of the linearly-dispersive Weyl cones. Opposite chiralities $\chi$ are associated with the two Weyl nodes. The Fermi level in undoped NbP is predicted to be above the intersection point of the Weyl cones, leading to a peanut-shaped cross-section of the Fermi surface of dimensions $k_{F1}$ and $k_{F1}$ and area $A_{FS}$. The Weyl Fermi surface is a fourfold symmetric representation of such cross-section. b) Simulated TEF spectrum for the Fermi surface as shown in a. c) Simulated TEF spectrum for the same Weyl Hamiltonian with a lower Fermi energy compared to b, for which the Weyl cones are separated. d) Experimental TEF spectrum for selected temperatures and E-C distance $L=$ 2.2 $\mu$m. Two prominent peaks are visible at $B=B_1$ and $B=B_2$. A smaller peak appears at $B=B_0$, consistently with the simulated spectrum in b. e) SEM micrograph of the NbP sample for TEF measurement after the milling of a hole to block long-traveler electrons (in orange) when $B=B_1$ (top) and let pass short-traveler electrons (in blue) when $B=B_2$ (bottom). f) TEF experiments with obstructed path and unobstructed path. The long-traveler electrons, which generate the first peak, are successfully blocked. g) Weyl nodes separation in NbP calculated using DFT packages Open-MX \cite{lee_fermi_2015} and VASP \cite{sun_topological_2015} compared with the measured value using HR-ARPES \cite{liu_evolution_2016} and the value obtained from TEF measurement (this work).}
\end{figure*}

\end{document}